\documentclass[prl,twocolumn,showpacs,preprintnumbers,amsmath,amssymb]{revtex4}

\usepackage{amsmath}

\usepackage{graphics}

\usepackage{epsf}

\usepackage{graphicx}
\usepackage{dcolumn}
\usepackage{bm}

\begin{document}

\title{Electron-induced stabilization of ferromagnetism in Ga$_{1-x}$Gd$_x$N}

\author{Gustavo M. Dalpian and Su-Huai Wei}
\affiliation{National Renewable Energy Laboratory, Golden, Colorado 80401, U.S.A}

\date{\today}

\begin{abstract}

Using {\it ab initio} band structure calculations and symmetry arguments, we show that the magnetic property of
Ga$_{1-x}$Gd$_x$N is drastically different from TM-doped GaN. The coupling between Gd atoms in the alloy
is antiferromagnetic, but the ferromagnetic phase can be stabilized by introducing electrons. Furthermore,
we propose a model that may explain the recently observed colossal magnetic moments in this system, based on the
polarization of donor electrons.

\end{abstract}

\pacs{75.50.Pp, 71.55.Eq, 71.70.-d, 71.20.Eh}

\maketitle

\section{introduction}

The development of future spintronic devices will likely rely heavily on
engineering materials that can have efficient spin injection into
semiconductors at high temperature \cite{zuticrmp,awsc00}. Among the most
promising candidates for this task are some diluted ferromagnetic
nitrides and oxides, \cite{diet00} which have small separation between
magnetic ions and small spin-orbit coupling at the band edge.  Most of the
materials that have been studied are semiconductors 
doped with partially filled $3d$ transition metals (TMs).  Because of the
strong coupling between the magnetic ions' 3$d$ states and the coupling to
the host $p$ states, diluted magnetic
nitride and oxide semiconductors have been predicted and, in some cases,
observed to show hole-mediated ferromagnetic behavior above room
temperatures \cite{zuticrmp,awsc00,diet00}. However, high $T_c$ behavior of the $3d$ 
TM-doped nitrides and oxides is often impeded by the formation of
precipitates \cite{dhar03} or compensated by $n$-type carriers that are intrinsic for
these wide gap semiconductors \cite{pankove,look99}. Recently, nitrides and oxides
doped with partially filled 4f rare-earth (RE) ions have been proposed
as an interesting alternative to achieve high $T_c$ in these
materials.  Indeed, ferromagnetism has already been observed in
Ga$_{1-x}$Gd$_x$N, with Curie temperature larger than 400\,K
\cite{teraguchi02,dhar05,asahi04}.

Despite the partial success, the nature of the host-impurity
couplings, and of the ferromagnetic interactions in Ga$_{1-x}$Gd$_x$N
and other RE-doped nitrides and oxides, is not very well understood.
On one hand, 4$f$ orbitals are more localized, thus the direct coupling
between the 4$f$ ions is expected to be week. On the other hand, 4$f$ RE
elements can have larger magnetic moments than the 3$d$ elements,
and, unlike the $d$ states, $f$ electrons can couple strongly with the
host $s$ electrons, leading to the possibility of electron-mediated
ferromagnetism in these materials. Furthermore, colossal magnetic
moments of Gd in GaN have been observed, which has a close connection
to the observed ferromagnetism in this system \cite{dhar05}. But the
origin of the observed colossal magnetic moments is not known.  

To get a deep understanding of the nature of magnetic coupling in these 
RE-doped systems, in this work, we perform total energy and band
structure calculations of Ga$_{1-x}$Gd$_x$N for a few concentrations,
including the hypothetical GdN in the zinc-blende structure. We show
that the interactions between Gd atoms in undoped Ga$_{1-x}$Gd$_x$N are
antiferromagnetic in nature, but this can be changed by introducing
donors to the material; different from most 3$d$ diluted magnetic
semiconductors, ferromagnetism in Ga$_{1-x}$Gd$_x$N is
electron-mediated \cite{coey}, because the splitting of the conduction
band, induced by the $s$-$f$ coupling, is larger.

\section{methodology}

The total energy and band structure calculations in this study were
performed using the general potential linearized augmented plane wave
(LAPW) method \cite{wei85} within the density functional theory and
the local spin density approximation (LSDA) for the
exchange-correlation potential.  The muffin-tin (MT) radii are 1.48
Bohr for N, 2.50 Bohr for Gd, and 2.05 Bohr for Ga.  The Gd $5p$ and Ga
$3d$ semicore states are treated in the same footing as the other
valence orbitals. The Brillouin-zone integrations were performed using
a 4x4x4 Monkhorst-Pack special $k$-points \cite{monk76} and equivalent
$k$-points for the superstructures. We find that the LSDA-calculated
lattice parameters for both GaN and GdN are in excellent agreement
with experiment.  Our calculated lattice parameters for GaN and GdN in the
zinc-blende structure are 4.497 and 5.291\,\AA. The difference in the lattice constant between
ferromagnetic (FM) and antiferromagnetic (AFM) configurations of GdN
is less than 0.2\%.  It should be noted that pure GdN usually
crystallizes in the NaCl structure.  We calculated the lattice
parameter of GdN in this structure, finding it to be 4.978\,\AA, which
is in excellent agreement with the experimental value of 4.974\,\AA
\cite{li97}.  For the alloys, we assume that Vegard's rule is obeyed
for the lattice constant, and the internal atomic positions are fully
relaxed by minimizing the total energy and quantum mechanical forces.
In this work we study Gd incorporated in zinc-blende GaN and assume
that the physical properties are similar for Gd incorporated in the more
stable wurtzite GaN.

\begin{figure}
\epsfxsize 8.5cm
\centerline{\epsffile{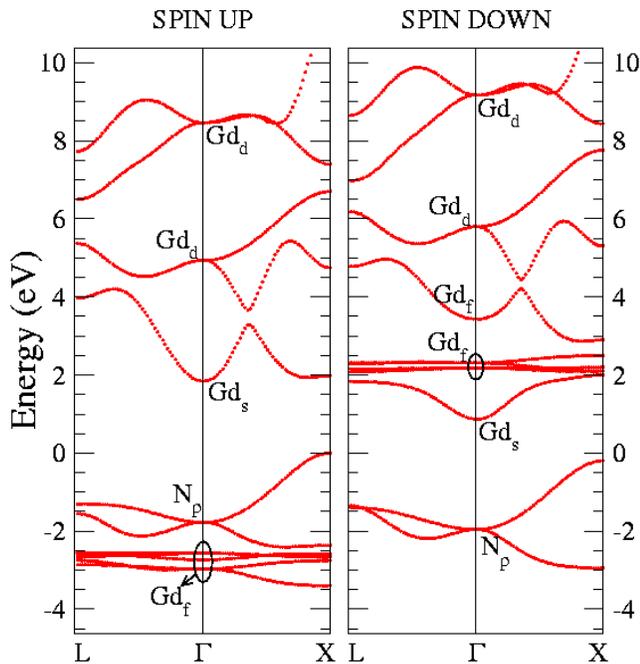}}
\caption{(Color online) Band structure of the ferromagnetic GdN in the hypothetical
zinc-blende structure. The labels
indicate the majority character of the band at the $\Gamma$ point.
\label{fig1}}
\end{figure}

\begin{figure}
\epsfxsize 8.5cm
\centerline{\epsffile{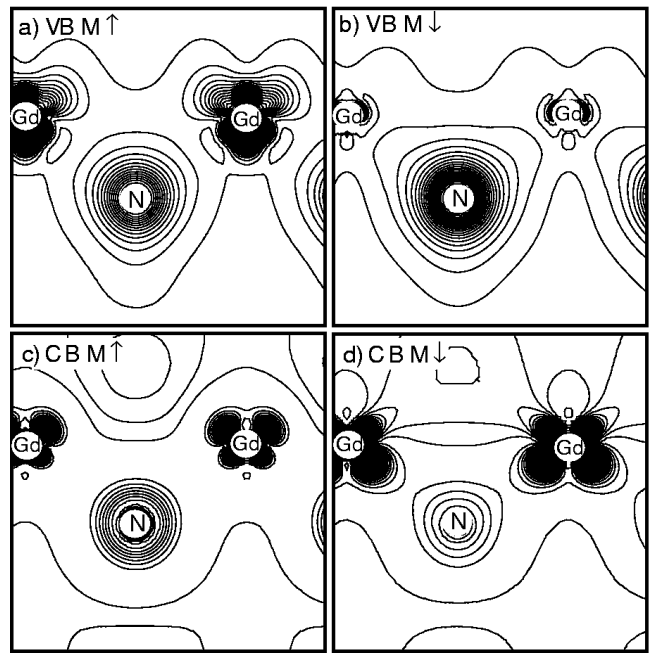}}
\caption{Wavefunction square of the (a) spin-up VBM, (b) spin-down VBM,
(c) spin-up CBM, and (d) spin-down CBM at the zone center.
\label{fig2}}
\end{figure}

\section{results and discussion}

To understand the effect of Gd incorporation in GaN, we first look at
the band structure (Fig. 1) of the hypothetical
zinc-blende ferromagnetic GdN compound and compare it with
that of MnN \cite{jano03}. We find that in the tetrahedral
environment, the crystal field splits the $f$ orbitals into $t_1$,
$t_2$, and $a_1$ states, which can couple to other $s$, $p$, and
$d$ states with the same crystal symmetry.  For Gd, the majority
spin-up $4f$ states are fully occupied and are below the valence band
maximum (VBM) state at $\Gamma$ with mostly N $2p$ character and
$t_{2}$ symmetry. The minority spin-down $4f$ states are fully unoccupied and are
above the conduction band minimum (CBM) at $\Gamma$. Unlike the $3d$
TM, the Gd $5d$ bands, with $e$ and $t_2$ symmetry, are high in energy,
and thus are fully unoccupied in the conduction band. The unique position
and characters of the Gd $4f$ and $5d$ orbitals lead to some unusual
features of GdN that are very different from those of Mn-doped GaN. (i)
The VBM state at $\Gamma$, with $t_{2p}$ character, couples strongly
with the Gd $t_{2d}$ and $t_{2f}$ states. In the spin-up channel, this
state is pushed up by the Gd $t_{2f}$ state below and pushed down by
the Gd $t_{2d}$ state above. In the spin-down channel, this state is
pushed down by both the Gd $t_{2f}$ and $t_{2d}$ states. However, the
$p$-$d$ coupling is larger in the spin-up channel, because the
potential exchange lowers the spin-up Gd $t_{2d}$ orbital
energy; the net effect is that the exchange splitting at the VBM ($N_0
\beta \sim -0.01$ eV) is much smaller than that observed in TM-doped
GaN.  Furthermore, because of the strong $p$-$d$ coupling at
$\Gamma$, which pushes the $t_{2p}$ state down, the top of the valence
band is not at $\Gamma$ as for MnN \cite{jano03}. (ii) Unlike in TM-doped
GaN, where $s$-$d$ coupling is not allowed in $T_d$ symmetry, $s$-$f$
coupling is allowed. Because the CBM at $\Gamma$ has the $s$
character, in the spin-up channel, it is pushed up by the occupied Gd
$a_{1f}$ state, and in the spin-down channel, it is pushed down by the
unoccupied Gd $a_{1f}$ state. This large kinetic exchange leads to a
large negative spin exchange splitting near the CBM ($N_0 \alpha \sim
-0.4$ eV). In contrast, the potential exchange induced splitting in 
TM-doped GaN is smaller and positive. The wavefunction square of the VBM and CBM 
states at $\Gamma$ are plotted in Fig. 2. We see that the spin-up VBM 
shows a large antibonding $p$-$f$ character. The $f$ character in the spin-down
VBM is relatively small because the Gd $f$ and VBM energy difference is 
large in the spin-down channel. For the CBM states, we find that the
spin-up CBM shows an antibonding $f$-$s$ character because the spin-up
$f$ state is below the CBM, whereas the spin-down CBM shows a larger bounding
$f$-$s$ character because the unoccupied spin-down $f$ level is above and
closer to the CBM. These results are consistent with the energy position 
and symmetry arguments above.

For diluted GaN:Gd magnetic semiconductors, the general features are
the same as in pure GdN.  The majority Gd $4f$ orbitals of Gd are
always inserted below the VBM of GaN, whereas the minority Gd $4f$
orbitals are always above the CBM. The Gd $5d$ orbitals, on the other
hand, always have higher energy, above the CBM of GaN. These features
can be seen in Figs. 3 and 4. In Fig. 3 we show the band structure for Ga$_{1-x}$Gd$_x$N with
$x=0.25$. 
Figure 4 shows the density of states (DOS) for Ga$_{1-x}$Gd$_x$N with
$x=0.0625$. We observe that the spin-up Gd $4f$ orbitals are resonant
within the valence band with large band widths, whereas the spin-down
Gd $4f$ orbitals are more localized.  We can also observe that the DOS
at the VBM is very high, indicating a very large valence effective
mass in this material.  The large
valence effective mass comes from the coupling between the VBM and the
Gd $d$ orbitals above, which can be seen in Fig. 4.
The repulsion between these orbitals pushes the
VBM at $\Gamma$ down, increasing the effective mass. This indicates
that this material will have a very small hole mobility.  On the other
hand, the DOS of the conduction band is very small. This can be
understood by the repulsion between the $4f$ levels and the CBM. This
interaction pushes the CBM at $\Gamma$ down in the spin-down channel,
decreasing the electron effective mass.  We find that, again, both
$N_0 \alpha$ and $N_0 \beta$ are negative for Ga$_{1-x}$Gd$_x$N, with
the magnitude of $N_0 \alpha$ much larger than that of $N_0 \beta$,
as can be seen in Fig. 3. 
This effect is an indication that, for Ga$_{1-x}$Gd$_x$N, the
stabilization of ferromagnetism will be more efficient when the
carriers are electrons instead of holes.

Because of the highly localized character of the $f$ orbitals, a natural
question is whether LSDA can successfully describe the
system. Fortunately Gd has a $4f^75d^16s^2$ valence configuration,
thus it is isovalent with Ga. The majority-spin $4f$ orbitals are
completely filled, whereas the minority-spin $4f$ orbitals are
completely empty. This situation is similar to Mn substituted II-VI diluted
semiconductors \cite{wei93}. The fact that the calculated lattice parameter for
GdN is in good agreement with experiment indicates that our LSDA
description is reasonable.  To further test the adequacy of the LSDA,
we have performed general gradient approximation (GGA) \cite{PW} and
model LDA+U type calculations \cite{wei93}. We find that besides small quantitative
changes (Table I), the qualitative picture described in this paper is unchanged. This can be understood
because GGA and LDA+U does not change the $f$-electron occupation in this system.

\begin{figure} 
\epsfxsize 8.5cm
\centerline{\epsffile{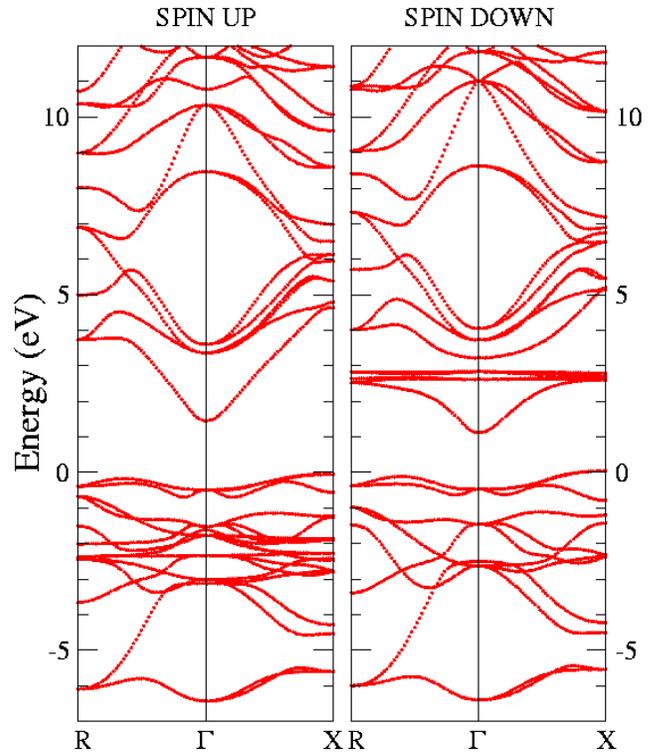}}
\caption{(Color online) Band structure for the ferromagnetic configuration of
Ga$_{1-x}$Gd$_x$N with $x=0.25$.
\label{fig3}}
\end{figure}

\begin{figure}
\epsfxsize 8.5cm
\centerline{\epsffile{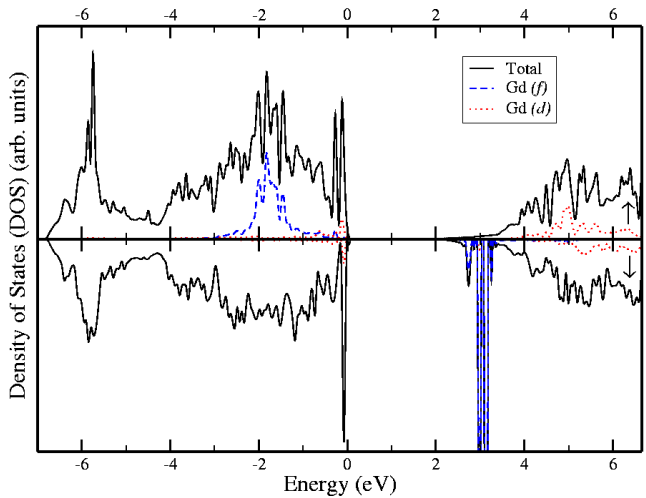}}
\caption{(Color online) Density of states of Ga$_{1-x}$Gd$_x$N with $x=0.0625$. The
blue (dashed) curves are partial DOS of Gd $f$ levels and the red (dotted) curves are partial DOS
of Gd$_d$. For clarification, the intensity of the DOS related to the Gd
$d$ levels is multiplied by a factor of three.
\label{fig4}}
\end{figure}

Our band structure calculation show that undoped zinc-blende GdN or
Ga$_{1-x}$Gd$_x$N alloys \cite{hashimoto03}, are semiconductors because
Gd is isovalent to Ga \cite{lambr00}. Our total energy calculations
find that undoped GdN and GaGdN in the zinc-blende phase are more
stable in the antiferromagnetic phase.  The calculated results
are shown in Table I for several Gd concentrations $x$ and
configurations.  There are always two Gd atoms in the unit cell.
These results are expected because no charge carriers
are created in this system, therefore, the spin-splitting at
the band edge present in the FM phase will not lead to any energy
gain compared to the unsplit band edge of the AFM phase. \cite{dalpian} On the
other hand, the superexchange coupling between the Gd $f$ states,
mediated by the N $p$ orbitals, stabilizes the AFM phase. However,
compared to TM-doped GaN, the energy differences ($\Delta E_{FM-AFM}$)
between the FM and AFM states for these systems are
small, which is due to the highly localized character of the $f$ orbitals.

In order to stabilize the FM phase, it is necessary to insert carriers
into the system.  As GaN is usually $n$-type doped with free electrons in the
conduction band \cite{pankove}, we have simulated the effect of donors
by introducing electrons into Ga$_{1-x}$Gd$_x$N.  In the AFM
configuration, the CBM of each spin channel have the same energy,
therefore, they will be occupied in both spin channels.  In the FM
configuration, because of the strong $s$-$f$ coupling, the CBM has negative
spin-exchange splitting, i.e., the spin-down CBM has a lower energy.
Consequently, when electrons are added, they will prefer to go to the spin-down 
CBM of the FM phase, stabilizing the FM configuration.  This expectation is
supported by our direct total energy calculations of charged systems,
where the FM phase becomes more stable. We have
considered two different configurations for $x=0.125$, where the Gd atoms are
nearest neighbors and next-nearest neighbors.  We find that the
effective exchange interaction between the two Gd [J$_{ij}$({\bf R})]
decreases as the distance between impurities becomes larger.

\begin{table}
\caption{Energy differences ($\Delta E_{FM-AFM}$) between ferromagnetic and
antiferromagnetic configurations in Ga$_{1-x}$Gd$_x$N. Energies are in meV; 
$nn$ and $nnn$ refer to the two Gd atoms as nearest and next-nearest
neighbors, respectively, in the unit cell.}
\begin{ruledtabular}
\begin{tabular}{ccc}
$x$ 		&	Charge (electrons/Gd)	&	$\Delta E_{FM-AFM}$	\\
1		&	0			&	25.0			\\
1               &       0                       &       22.0   (GGA)                 \\
0.25		&	0			&	2.6			\\
0.125 (nn)	&	0			&	4.4			\\
0.125 (nn)	&	0.1			&	-28.4			\\
0.125 (nnn)	&	0			&	1.3			\\
0.125 (nnn)	&	0.1			&	-19.5			\\
\label{table1}
\end{tabular}
\end{ruledtabular}
\end{table}

Several experimental studies \cite{teraguchi02,dhar05,asahi04}
of Ga$_{1-x}$Gd$_x$N claim that
ferromagnetism is observed up to room temperature.  Our
calculations show that these observed FM phases should be related to the $n$-type
character of the Ga$_{1-x}$Gd$_x$N samples.  This is supported
by photoluminescence measurement, which shows the presence of large
defect concentration in the sample with FM ordering \cite{zhou04}.  A recent 
measurement also shows that the presence of FM is correlated with Oxygen, a known 
donor defect in the sample \cite{dhar05}, with a concentration of
about $10^{18}/cm^3$. These observations are consistent with our model.

One of the unique features observed in Gd-doped GaN is the colossal
magnetic moment \cite{dhar05}. It was observed that, at very low Gd
concentrations ($\sim 10^{16}/cm^3$), the total effective magnetic
moment per Gd atom could be as high as 4000 $\mu_B$. This is quite
unusual because the atomic moment of Gd is only 8 $\mu_B$, and in the
hypothetical zinc-blend GdN the moment per Gd is only 7 $\mu_B$.
When the Gd concentration increases, the effective moments per Gd
decrease. This observation was explained through a phenomenological
model, which assumes that Gd atoms polarize the GaN matrix in a
certain radius around them and each host atom within the radius has an
induced magnetic moment $p_0$. Although this model is very
interesting and fits experimental results quite well, at least in
the low-concentration region, it does not provide a mechanism of the
induced magnetic moment. Our band structure calculations above,
however, provide a possible explanation of the observed huge magnetic
moments: When Gd is introduced into GaN, each Gd introduces unoccupied
$f$ states above the spin-down CBM. The symmetry-allowed coupling
between the $f$ states and states below it can create localized levels
with energy below the original CBM.  When the system also contains
enough shallow donor impurities, as reported in the experiment, two
things will occur: (i) the donor electrons of the impurities will
occupy the Gd-induced empty levels, if the donor levels of the
impurities are higher than the Gd-induced level;  (ii)
consequently, the system will become ferromagnetic. When enough levels
are created by Gd below the donor level, which is possibly due to the
symmetry-allowed $s$-$f$ coupling, the induced magnetic moment per Gd
atom can be very high in the extreme impurity limit when the Gd-Gd
distances are very large.  Translating this into a real-space
representation, it is equivalent to say that atoms with a certain
radius $r_0$ around each Gd atom has its spin-down energy levels below
the donor levels, and thus they are polarized by the Gd atom with
opposite spins. As Gd concentration increases, the spheres surrounding each Gd atom
start to overlap.  Although some more levels could be pushed below
the donor level, thus increasing slightly the effective radius $r_0$, eventually the
overlap will reduce the effective volume affected by each individual Gd,
and therefore, reduce the magnetic moment per Gd, as observed in
experiment. However, unlike the previously proposed model, our model has two
distinct physical characteristics: (i) To have a large induced magnetic
moment, the system must have enough donor impurities; (ii) Our model suggests that the
induced magnetic moment has the opposite sign of the Gd magnetic
moment. This indicates that when the Gd concentration increases, or the
donor carrier density decreases, the magnetic moments in
the system will change sign. Experimental testing of our model is called
for.

It is interesting to point out that this electron-induced stabilization of
ferromagnetic order and large magnetic moments may also occur in TM-doped
semiconductors \cite{coey}. This is because, despite that $s$-$d$ coupling is
not allowed under $T_d$ symmetry, it will be allowed when the symmetry of 
the system is reduced either by strain or by chemical ordering. For example,
under (001) strain, the $a_1(s)$ state can couple to the split $a_1(e_d)$
state, whereas under (111) strain or in wurtzite crystal, the $a_1(s)$ state 
can couple to the split $a_1(t_d)$ and $a_1(t_p)$ state. However, in general, 
the $s$-$d$ coupling is expected to be smaller than the symmetry-allowed $s$-$f$ 
coupling in these tetrahedral semiconductor systems.

\section{summary}

In summary, we have investigated in detail the electronic and
magnetic properties of Ga$_{1-x}$Gd$_x$N.  We find that because of the
coupling between the Gd $f$ and host $s$ states, this system shows some
unique behavior that is drastically different from TM-doped GaN. The exchange splitting at the
CBM is negative. The coupling between Gd atoms is found to be antiferromagnetic, 
if no donors are present, but it become ferromagnetic when enough 
donors are present in the system. We also proposed a model that may explain the
colossal magnetic moments observed in this system in the very dilute limit, 
showing that it should be directly related to the polarization of donor electrons.

\section{acknowledgments}
The work at NREL is funded by the U.S. Department of Energy, Office of
Science, Basic Energy Sciences, under Contract No. DE-AC36-99GO10337
to NREL.

\end{document}